\documentclass{pasj00}
\draft

\begin{document}
\SetRunningHead{A.Tanikawa and T.Fukushige}{Mass Loss Timescale of Star Clusters in External Tidal Field}
\title{Mass Loss Timescale of Star Clusters \\
in External Tidal Field}
\author{Ataru \textsc{Tanikawa}, and Toshiyuki \textsc{Fukushige}}
\affil{Department of General System Studies, College of Arts and Sciences,\\
University of Tokyo, 3-8-1 Komaba, Meguro-ku, Tokyo 153-8902}
\email{tanikawa@provence.c.u-tokyo.ac.jp}
\maketitle

\begin{abstract}
We investigate evolution of star clusters in external tidal field by means of
$N$-body simulations.
We followed seven sets of cluster models whose central concentration and
strength of the tidal field are different.
We found that the mass loss timescale due to escape of stars, $t_{\rm mloss}$,
and its dependence on the two-body relaxation timescale, $t_{\rm rh,i}$, are
determined by the strength of the tidal field.
The logarithmic slope [$\equiv d\ln(t_{\rm mloss})/d\ln(t_{\rm rh,i})$]
approaches to near unity for the cluster models in weaker tidal field.
The timescale and the dependence are almost independent of the central
concentration for clusters in the tidal field of the same strength.
In our results, the scaling found by \citet{Baum2001} can be seen only in the
cluster models with moderately strong tidal field.
\end{abstract}

\section{Introduction}
The escape of stars from clusters is a long-standing problem in the study of
dynamical evolution of star clusters.
The theory for the escape was first developed by \citet{Ambart} and
\citet{Spit1940}.
It is based on the assumption that distant encounters between cluster stars
will gradually set up a Maxwellian velocity distribution.
Some stars have escape velocity of the cluster and they escape.
The timescale that the distant encounters set up a Maxwellian velocity
distribution corresponds to the two-body relaxation timescale (\cite{Chandra},
\cite{Spit1987}):
\begin{equation}
t_{\rm r} = 0.065 \frac{{v_{\rm m}}^3}{n m^2 G^2 \ln \Lambda},
\end{equation}
where $n$ is the number density of stars, $m$ is the mass of the cluster stars,
$v_{\rm m}$ is the average velocity of the stars, $G$ is the gravitational
constant, and $\Lambda$ is the Coulomb logarithm.
Every time the two-body relaxation time elapses, a cluster loses constant
fraction of stars which have escape velocity.
Therefore, the mass loss timescale of a cluster is proportional to the two-body
relaxation time.

This theoretical argument was confirmed by \citet{Baum2001} (hereafter, B01)
in the absence of the external tidal field.
He performed $N$-body simulations of evolution of star clusters, and showed
that the mass loss of the cluster happened on the two-body relaxation
timescale for energy cutoff clusters, and on the two-body relaxation with
backscattering correction for radial (spatial) cutoff clusters.

However, the mass loss timescale from the clusters in external tidal field is
remained unclear.
The Collaborative Experiments \citep{Heggie1998} demonstrated that the mass
loss timescale does not scale with the two-body relaxation timescale, where
multi-mass clusters moving in a steady tidal field were simulated.
It turns out that the complication is due to the existence, at the initial
setup, of population of stars that have energies above the escape energy and
remain in a cluster before finding exits (''potential escapers'').
\citet{Fuku2000} (hereafter, FH) quantified the escape timescale of
the potential escaper as a function of energy, and pointed out the escape
timescale of the potential escaper is long enough to influence the whole mass
loss timescale of clusters.

B01 performed $N$-body simulations of star clusters, where
equal-mass clusters of $W_{0}=3$ King profiles \citep{King1966} evolve in the
steady external tidal field.
He investigated the dependence of mass loss timescale on two-body relaxation
times using simulations whose particle numbers are $N=128$ to $16384$.
He found that the mass loss timescale of clusters, $t_{\rm mloss}$, are
proportional to, ${t_{\rm rh}}^{3/4}$, the half-mass relaxation time to
the power of $3/4$, where the half-mass relaxation time is given by
\begin{equation}
t_{\rm rh} = 0.138 \frac{\sqrt{N} {r_{\rm h}}^{3/2}}{\sqrt{m} \sqrt{G} \ln(\gamma N)},
\end{equation}
\citep{Spit1987} where $N$ is the number of stars, $r_{\rm h}$ is the
half-mass radius, and $\gamma=0.11$ \citep{Gier1994}.
\citet{Baum2003} found almost the same scaling law for multi-mass clusters in
time-dependent tidal field.

However, the $t_{\rm mloss} \propto {t_{\rm rh}}^{3/4}$ scaling
is not convinced as an asymptotic behavior at larger $N$ (or $t_{\rm rh}$)
for the following reason.
In the limit of large $N$, the two-body relaxation timescale,
$t_{\rm rh}$, is much longer than the escape time delay,
$t_{\rm e}$, which is the duration from the moment when the star gets
energy above the escape energy to that when it actually escapes from the
cluster.
Therefore, the mass loss timescale should be determined only by the half-mass
relaxation time, i.e. $t_{\rm mloss} \propto t_{\rm rh}$.

The purpose of this paper is to investigate whether the
$t_{\rm mloss} \propto {t_{\rm rh}}^{3/4}$ scaling obtained by
B01 is an asymptotic behavior by means of $N$-body simulations.
We perform sets of simulations of star clusters whose two-body relaxation
timescale is larger than that of B01 and that are in the external
tidal field with different strength.

The structure of the paper is as follows.
We describe in detail the model of the cluster in section 2.
We show the results of $N$-body simulations in section 3.
section 4 is for discussion.
In section 5, we summarize this paper.

\section{Simulation Method}
We investigate the evolution of star clusters in the external tidal field by
means of $N$-body simulations.
We adopt the conventional model in which the cluster is assumed to
move on a circular orbit in a spherically symmetric galaxy potential,
taken to be that of a distant point mass $M_{\rm g}$.
We set the initial center of mass of the cluster at the origin
$(x, y, z) = (0, 0, 0)$, with axes oriented so that the position
of the galactic center is $(-R_{\rm g}, 0, 0)$.
We assume that the size of the globular cluster is much smaller than
$R_{\rm g}$.
If the globular cluster rotates around the galactic center at an angular
velocity ${\boldsymbol{\Omega}} = (0, 0, \omega)$, the equation of
motion of an individual cluster member can be expressed as
\begin{equation}
\frac{d^2{\boldsymbol{r}}_{i}}{dt^2} = - \nabla \Phi_{{\rm c},i} - 2 {\boldsymbol{\Omega}} \times \frac{d {\boldsymbol{r}}_{i}}{dt} + \omega^2 (3 x_{i} {\boldsymbol{\hat{e}}}_{\rm x} - z_{i} {\boldsymbol{\hat{e}}}_{\rm z}), \label{eq:motion}
\end{equation}
where ${\boldsymbol{r}}_{i}$ is the position of the $i$-th particle,
and ${\boldsymbol{\hat{e}}}_{\rm x}$, ${\boldsymbol{\hat{e}}}_{\rm z}$
are unit vectors that point along the $x$, $z$ axes, respectively.
The first term on the right-hand side in equation (\ref{eq:motion}) is
the gravitational acceleration from other particles in the cluster,
the second term is the Coriolis acceleration, and the third term
is a combination of the centrifugal and tidal forces.
We used standard units \citep{Heggie1986}, such that
$M_{\rm i}=G=-4E_{\rm c}=1$, where $M_{\rm i}$ is the
initial total mass and $E_{\rm c}$ is the initial total energy within
the cluster.
In our simulations we used a softened gravitational potential expressed as
\begin{equation}
\displaystyle - \nabla \Phi_{{\rm c},i} = - \sum^{N}_{j=1,j \neq i} \frac{G m_{j} ({\boldsymbol{r}}_{i} - {\boldsymbol{r}}_{j})}{({|{\boldsymbol{r}}_{i} - {\boldsymbol{r}}_{j}|}^2 + \varepsilon^2)^{3/2}}, \label{eq:phici}
\end{equation}
where $m_{j}$ is mass of $j$-th particle and $\varepsilon$ is a softening
parameter.
We set the softening parameter, $\varepsilon$, as $1/32$.

\begin{table*}
\begin{center}
\caption{Initial cluster models}
\begin{tabular}{ccccccc}
\hline
\hline
Model name & $W_{0}$ & $r_{\rm t,i}$ & $r_{\rm t,i}/r_{\rm kg}(W_{0}=3)$ & $r_{\rm t,i}/r_{\rm kg}(W_{0}=7)$ & $N$ & $t_{\rm dy,r=r_{t,i}}$ \\
\hline
k3r0.8     & $3$ & $2.50$ & $0.80$ & $0.36$ & $128 - 131072$ & $7.9$ \\
k3r1.0     & $3$ & $3.13$ & $1.0$ & $0.44$ & $128 - 131072$ & $11$ \\
k3r1.3     & $3$ & $4.20$ & $1.3$ & $0.60$ & $128 - 32768$ & $17$ \\
k3r2.2     & $3$ & $6.98$ & $2.2$ & $1.0$ & $128 - 32768$ & $37$ \\
k3r4.5      & $3$ & $14.0$ & $4.5$ & $2.0$ & $128 - 16384$ & $100$ \\
\hline
k7r1.0     & $7$ & $3.13$ & $1.0$ & $0.44$ & $128 -131072$ & $11$ \\
k7r2.2     & $7$ & $6.98$ & $2.2$ & $1.0$ & $128 - 32768$ & $37$ \\
\hline
\hline
\end{tabular}
\label{tab:series}
\end{center}
\end{table*}

\begin{figure}
\begin{center}
\FigureFile(120mm,75mm){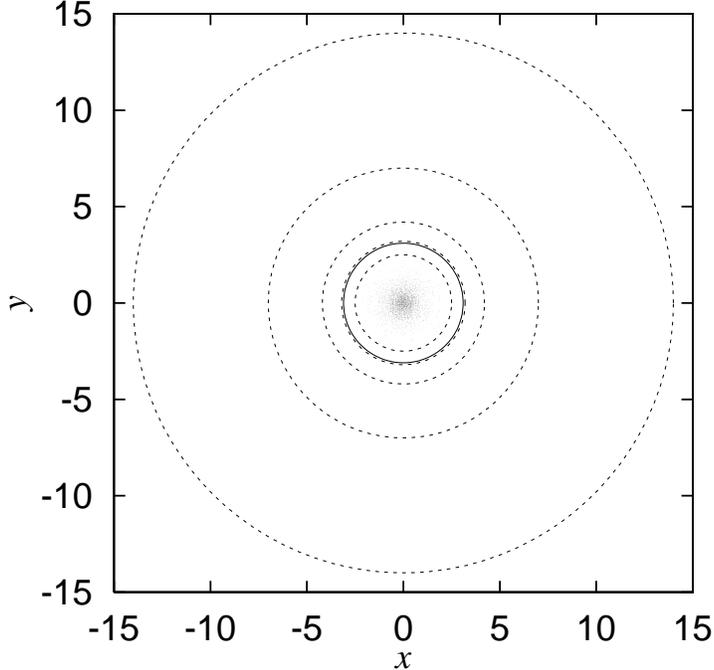}
\end{center}
\caption{Tidal radii, $r_{\rm t,i}$, together with the particle
distribution ($N=4096$), for models with $W_{0}=3$ King model.
The dashed circles show the tidal radii ($r_{\rm t,i}$) determined by the
external tidal field outward for models k3r0.8, k3r1.0, k3r1.3, k3r2.2 and
k3r4.5, and the solid circle shows the King cutoff radius ($r_{\rm kg}$)
beyond which the density is zero in King model.}
\label{fig:fig1}
\end{figure}

We used King's models \citep{King1966} to generate initial distribution of
star clusters.
We perform seven sets of simulations of star clusters whose initial
dimensionless central potential, $W_{0}$, of King model and initial tidal
radii determined by the external field, $r_{\rm t,i}$ are different,
which are summarized in table 1.
The number after ''k'' in model name indicates $W_{0}$ of King model.
We set the strength of the external tidal field by giving the angular velocity
$\omega$ in equation (\ref{eq:phici}) using a relation,
$\omega = \sqrt{G M_{\rm i} / 3 {r_{\rm t,i}}^3}$.
The number after ''r'' in the model name indicates the tidal radius scaled by
$r_{\rm kg}$ of $W_{0}=3$, where $r_{\rm kg}$ is the radius
beyond which the density is zero in King model (hereafter, King cutoff radius).
The values for $r_{\rm kg}$ in the standard unit are $3.13$ and $6.98$
for $W_{0}=3$ and $7$, respectively.
Figure \ref{fig:fig1} illustrates the tidal radii, $r_{\rm t,i}$
for simulations of $W_{0}=3$ King model.
The number of particles used for runs are listed in table 1.
All particles have the same mass, $m=M_{\rm i}/N$.
We perform five runs whose realization of particle distribution are different
for each $N$ when $N \leqq 8192$, and one run which $N \geqq 16384$.

We perform numerical integrations of equation (\ref{eq:motion}) using a leap-frog
integration scheme with shared and constant timestep.
The stepsize, $\Delta t$, is set to be as $1/64$ in models k3r0.8, k3r1.0 and
k7r1.0, and as $1/128$ in models k3r1.3, k3r2.2, k3r4.5 and k7r2.2.

The force calculation is done with the Barnes-Hut algorithm
\citep{Barnes1986} on GRAPE-5 \citep{Kawai2000}, a special-purpose computer
designed to accelerate $N$-body simulations.
Actually, we used the same code as in \citet{Fuku2001}.
We use only the dipole expansion and the opening parameter $\theta = 0.5$.
It took about 500 CPU hours to complete the most time-consuming run,
$N=32768$ of model k3r2.2 (about $7.2 \times 10^6$ timesteps).
For smaller $N$ simulations, the force calculation is done by direct
summations (when $N \leqq 4096$) and on host computer (without GRAPE-5, when
$N \leqq 512$).

Contrary to the standard star cluster simulations, our simulation uses the
softened gravitational potential and the leap-flog integrator with relatively
large stepsize.
And also, the force calculation is done with the tree algorithm.
These approaches are adopted in order to make the two-body relaxation timescale
as large as possible.
As will be discussed in the appendix, the approaches we adopted do not
influence the results concerning on the escape from the cluster.

\section{Results}
\subsection{Evolution of Total Mass}
\begin{figure*}
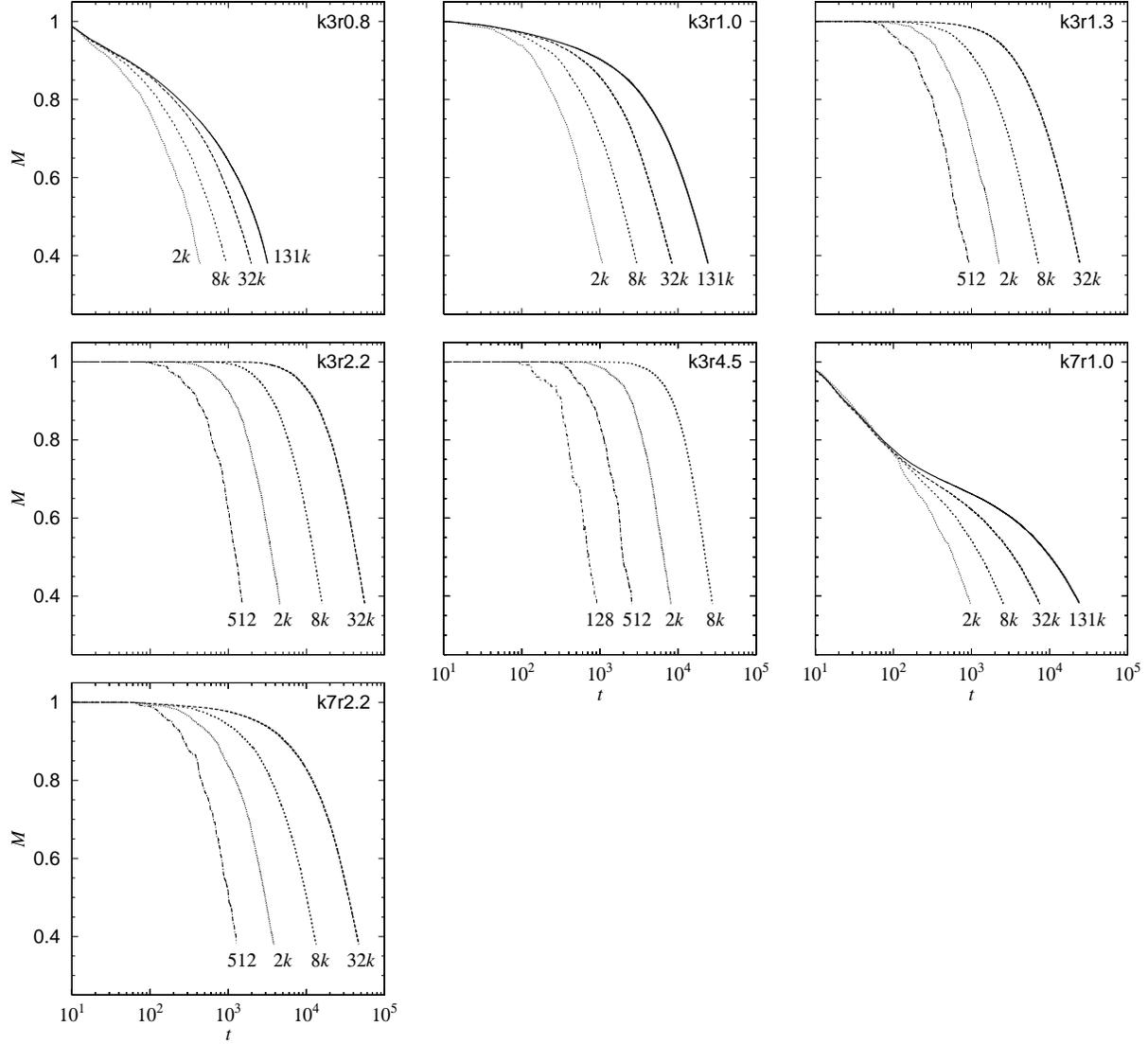

\begin{center}
\FigureFile(160mm,100mm){figure2.eps}
\end{center}
\caption{Evolutions of the mass of the clusters.
The numbers under the curves indicate the numbers of particles, $N$.}
\label{fig:fig2}
\end{figure*}

Figure \ref{fig:fig2} shows evolution of the total mass for all cluster
models.
The curves indicate the decrease in mass of the cluster defined by a tidal
boundary.
We define geometrically the cluster member as all stars within the tidal
radius from the center of mass of the cluster.
The tidal radius is expressed as
\begin{equation}
r_{\rm t} = \left( \frac{G M}{3 \omega^2}\right)^{1/3},
\end{equation}
where $M$ is total mass of the members of the cluster at a given time.
Since $M$ depends on $r_{\rm t}$ itself, some iteration is usually
required.
We remove stars when they escape far enough (more than 4096 from the coordinate
origin).

\subsection{Mass Loss Timescale}
\begin{figure}
\begin{center}
\FigureFile(120mm,75mm){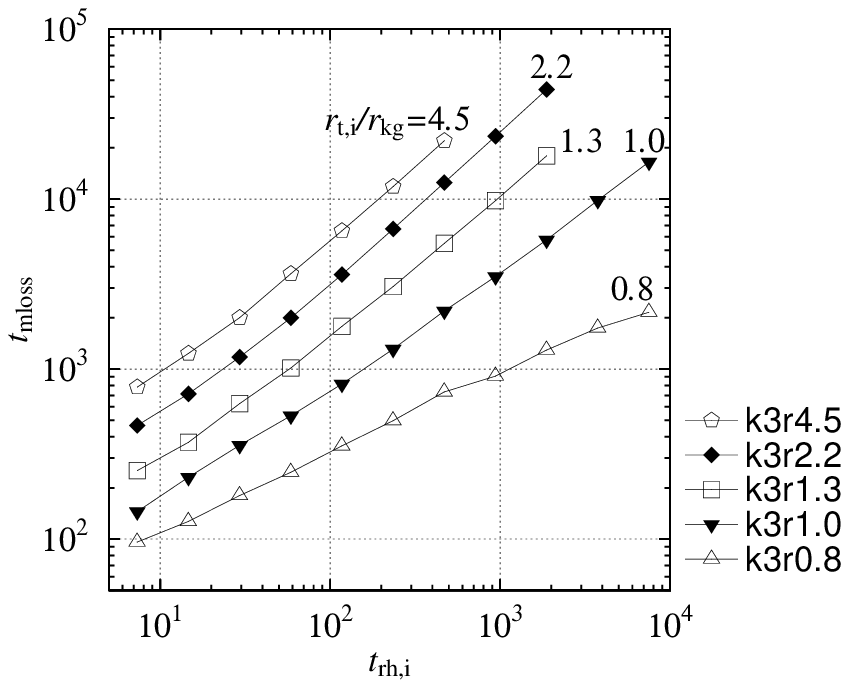}
\end{center}
\caption{Mass loss timescale of the clusters as a function of initial half-mass
two-body relaxation timescale for $W_{0}=3$ King cluster models.
The numbers near the curves indicate $r_{\rm t,i}/r_{\rm kg}$.}
\label{fig3}
\end{figure}

Figure \ref{fig3} shows the mass loss timescale,
$t_{\rm mloss}$, of clusters as a function of the initial half-mass
relaxation time, $t_{\rm rh,i}$, for $W_{0}=3$ King cluster models.
The mass loss timescale is here defined as the time when 50 \% of the initial
total mass is lost.
Since we perform five runs when $N \leqq 8192$, we adopt the means of these
runs as the mass loss timescale, $t_{\rm mloss}$.
Table \ref{tab:devi} shows the maximum deviations among these runs.
The maximum deviation is defined as the largest difference from the mean.
Since we use the potential softening and fix the softening parameters for all
models, the initial half-mass relaxation time, $t_{\rm rh,i}$, is
expressed as
\begin{equation}
t_{\rm rh,i} = 0.138 \frac{N {r_{\rm h,i}}^{3/2}}{{M_{\rm i}}^{1/2} G^{1/2} \ln (0.25 r_{\rm h,i} / \varepsilon)}, \label{eq:smoothed}
\end{equation}
where $r_{\rm h,i}$ is the initial half-mass radius and
$\varepsilon = 1/32$.
Thus, the initial half-mass relaxation timescale, $t_{\rm rh,i}$, is
estimated as $t_{\rm rh,i} = 470 \times (N/8192)$.
The initial half-mass radii in the standard unit are $r_{\rm h,i}=0.84$.

\begin{table*}
\begin{center}
\caption{Maximum deviation of $t_{\rm mloss}$ for $N \leqq 8192$.}
\begin{tabular}{c|ccccc|cc}
\hline
\hline
N & k3r0.8 & k3r1.0 & k3r1.3 & k3r2.2 & k3r4.5 & k7r1.0 & k7r2.2 \\
\hline
128 & 11 & 32 & 62 & 40 & 86 & 41 & 151 \\
256 & 21 & 20 & 41 & 65 & 145 & 33 & 94 \\
512 & 22 & 19 & 40 & 90 & 129 & 48 & 53 \\
1024 & 50 & 108 & 40 & 38 & 189 & 57 & 121 \\
2048 & 37 & 40 & 115 & 230 & 138 & 36 & 123 \\
4096 & 76 & 81 & 145 & 194 & 419 & 51 & 183 \\
8192 & 92 & 166 & 201 & 118 & 286 & 200 & 211 \\
\hline
\hline
\end{tabular}
\label{tab:devi}
\end{center}
\end{table*}

\begin{figure}
\begin{center}
\FigureFile(120mm,75mm){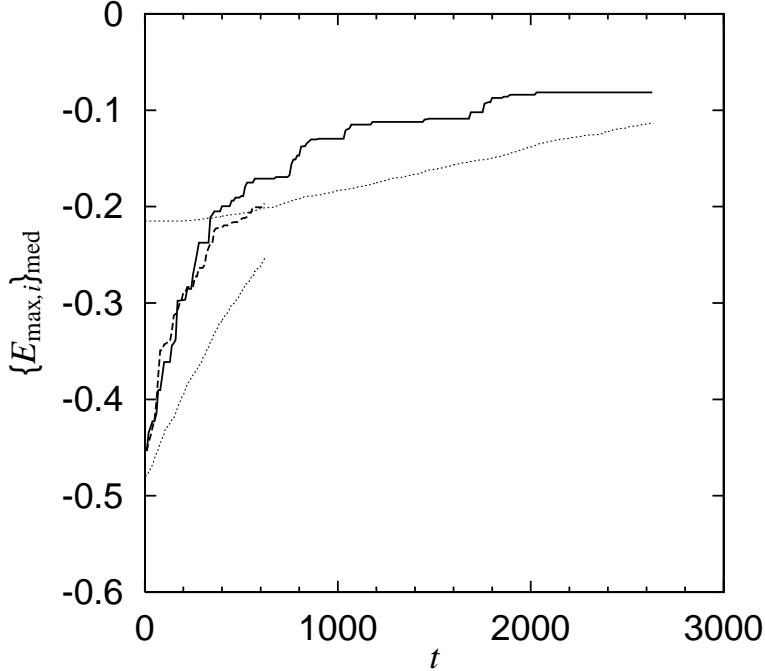}
\end{center}
\caption{Change in energy of an individual star for k3r1.0 and k3r2.2 models
($N=1024$).
The dashed and solid lines indicate
$\{E_{{\rm max},i}\}_{{\rm med}}$ (the definition is in the text) in
models k3r1.0 and k3r2.2, respectively.
The upper dotted line indicates the escape energy in model k3r2.2 and the
lower dotted line indicates that in model k3r1.0.}
\label{fig:fig4}
\end{figure}

As can be seen from figure \ref{fig3}, a cluster in stronger tidal
field lose its mass sooner.
The difference of the mass loss timescale is considered to be due to the
difference of the escape energy (the potential height at the Lagrangian point),
which is lower for a cluster in stronger tidal field.
Figure \ref{fig:fig4} shows change in energy of an individual star for k3r1.0
and k3r2.2 models ($N=1024$).
In figure \ref{fig:fig4} we plot $\{E_{{\rm max},i}\}_{\rm med}$,
the median value of the maximum energy records at a given time, $t_{1}$;
\begin{equation}
E_{{\rm max},i}(t_{1}) = \max_{t<t_{1}} \{ {E_{i}(t)} \},
\end{equation}
\begin{equation}
E_{i}(t) = \frac{1}{2} ({v_{{\rm x}i}}^2 + {v_{{\rm y}i}}^2 + {v_{{\rm z}i}}^2) + \Phi_{{\rm c},i} + (-\frac{3}{2} {x_{i}}^2 + \frac{1}{2} {z_{i}}^2)
\end{equation}
of randomly selected 50 particles, where $v_{{\rm x}i}$,
$v_{{\rm y}i}$, and $v_{{\rm z}i}$ are ${\rm x}$,
${\rm y}$, and ${\rm z}$ components of velocity of $i$-th particle,
respectively.
We do not update $E_{{\rm max},i}(t_{1})$ after stars escape from the
cluster.
The representative value, $\{E_{{\rm max},i}\}_{\rm med}$, may
trace the energy acquired by two-body relaxations, on average.
When stars obtain enough energy, they escape from cluster eventually.
The dashed lines indicate the escape energy of each models, expressed as
\begin{equation}
E_{\rm crit} = - \frac{3}{2} \frac{GM}{r_{\rm t}}.
\end{equation}

As shown in figure \ref{fig:fig4}, the increase of
$\{E_{{\rm max},i}\}_{{\rm med}}$ at $t<600$ is very similar among
models k3r1.0 and k3r2.2, which indicates that the relaxation processes in both
models occur with almost the same timescale.
However, because the escape energy in weaker tidal field is larger, it takes
more time for a significant fraction of stars to escape in model k7r2.2 than in
model k7r1.0.

\begin{figure}
\begin{center}
\FigureFile(120mm,75mm){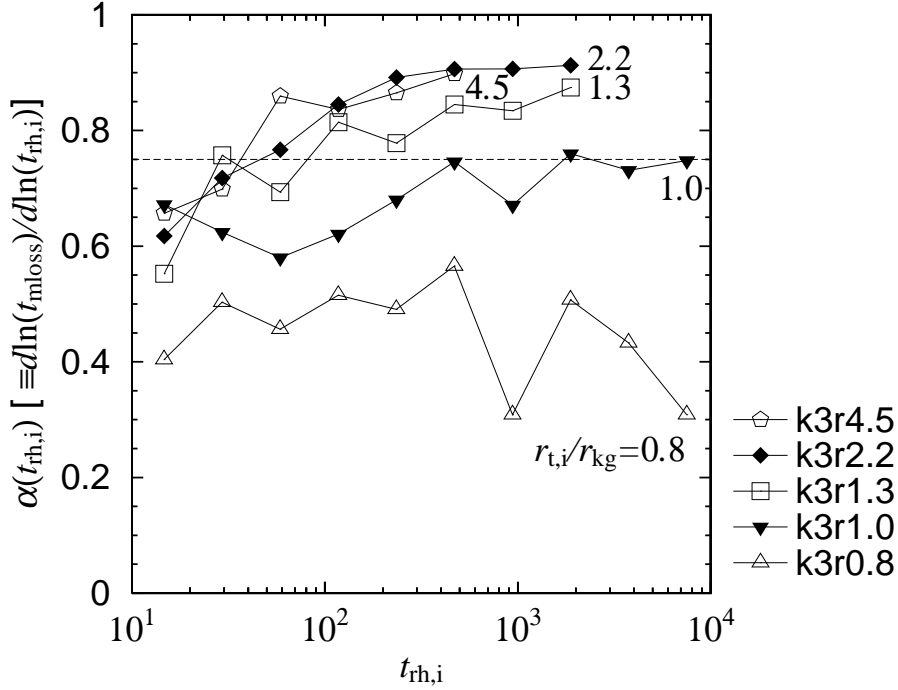}
\end{center}
\caption{Logarithmic slope $\alpha (t_{\rm rh,i})$ [ $\equiv d \ln (t_{\rm mloss}) / d \ln (t_{\rm rh,i})$]
of dependence on the initial half-mass relaxation timescale.
The dashed line indicates $\alpha (t_{\rm rh,i})= 0.75$.}
\label{fig:fig5}
\end{figure}

Figure \ref{fig:fig5} shows logarithmic slopes,
$\alpha (t_{\rm rh,i}) \equiv d \ln ({t_{\rm mloss}}) / d \ln (t_{\rm rh,i})$,
of the dependence on $t_{\rm rh,i}$ of the mass loss timescale shown in
figure \ref{fig3}.
Actually, we calculated the logarithmic slope $\alpha (t_{\rm rh,i})$
using a relation;
\begin{equation}
\alpha (t_{\rm rh,i}) = \frac{\log_{10}[t_{\rm mloss} (t_{\rm rh,i}) / t_{\rm mloss} (t_{\rm rh,i}/2)]}{\log_{10}[t_{\rm rh,i}/(t_{\rm rh,i}/2)]},
\end{equation}
where $t_{\rm mloss} (t_{\rm rh,i})$ is the mass loss timescale when
the initial half-mass relaxation timescale is $t_{\rm rh,i}$.

As can be seen in figure \ref{fig:fig5}, the logarithmic slope depends on
the the strength of external tidal field.
In model k3r1.0, where the initial distribution ($W_{0}=3$ King model) and
the tidal field ($r_{\rm t,i} = r_{\rm kg}$) are set to be the
same as in B01, we reproduce the asymptotic power,
$\alpha (t_{\rm rh,i}) \sim 0.75$, obtained by him.
Note that the initial half-mass relaxation timescale of the run with the
largest $N$ in B01 is about $230$.
On the other hands, when the tidal field is stronger (model k3r0.8) the slope
is smaller [$\alpha (t_{\rm rh,i}) \sim 0.4$], and when it is weaker
(models k3r1.3, k3r2.2, and k3r4.5) the slope is larger than
$\alpha (t_{\rm rh,i}) = 0.75$.
The slope $\alpha (t_{\rm rh,i})$ for the models in much weaker tidal
field (more k3r2.2 and k3r4.5) seems to approach asymptotically
$\alpha (t_{\rm rh,i}) \sim 0.9$, not $1$.

\begin{figure*}
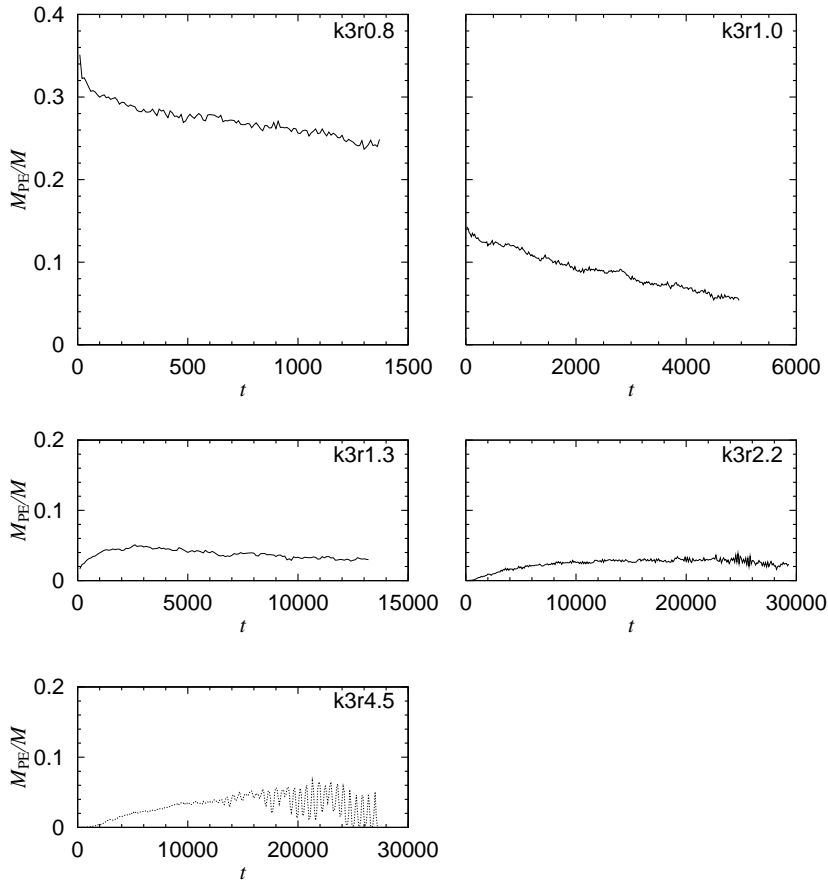

\begin{center}
\FigureFile(110mm,70mm){figure6.eps}
\end{center}
\caption{Evolution of the fraction of the mass of potential escapers in the
King $W_{0}=3$ clusters.}
\label{fig:fig6}
\end{figure*}

The dependence of the slope is considered to be associated with the population
of the potential escaper.
For the clusters whose initial half-mass relaxation timescale is smaller (or
smaller $N$), the escape time delay due to the potential escaper delays more
the total mass loss of clusters.
Figure \ref{fig:fig6} shows the evolution of the fraction of
potential escapers $M_{\rm pe}/M$ in the King $W_{0}=3$ cluster models
($N=16384$, except for model k3r4.5, $N=8192$).
We can see that as the slope $\alpha (t_{\rm rh,i}) $ shown in
Figure \ref{fig:fig5} decreases, the fraction $M_{\rm pe}/M$
increases.

\subsection{Effect of Initial Concentration}
\begin{figure}
\begin{center}
\FigureFile(120mm,75mm){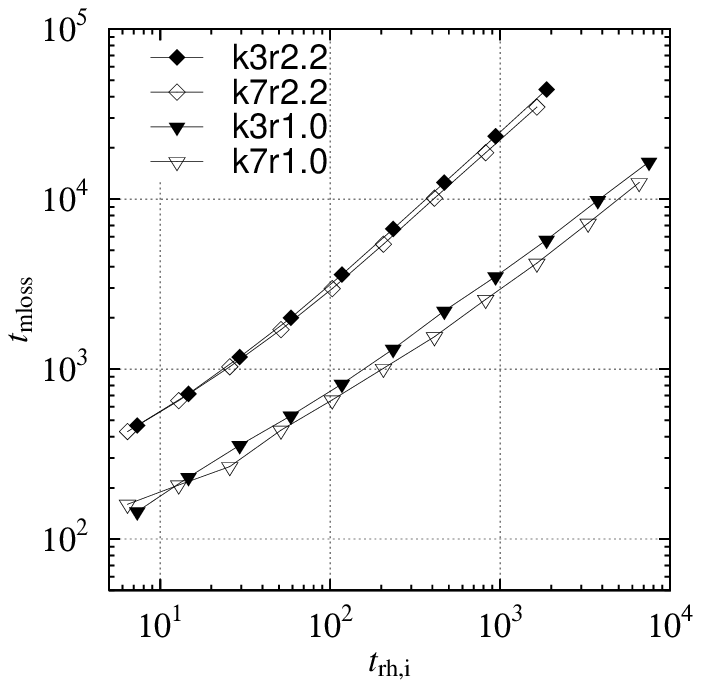}
\end{center}
\caption{Mass loss timescale as a function of the initial half-mass relaxation
timescale for models k3r1.0, k3r2.2, k7r1.0 and k7r2.2.}
\label{fig:fig7}
\end{figure}

\begin{figure}
\begin{center}
\FigureFile(120mm,75mm){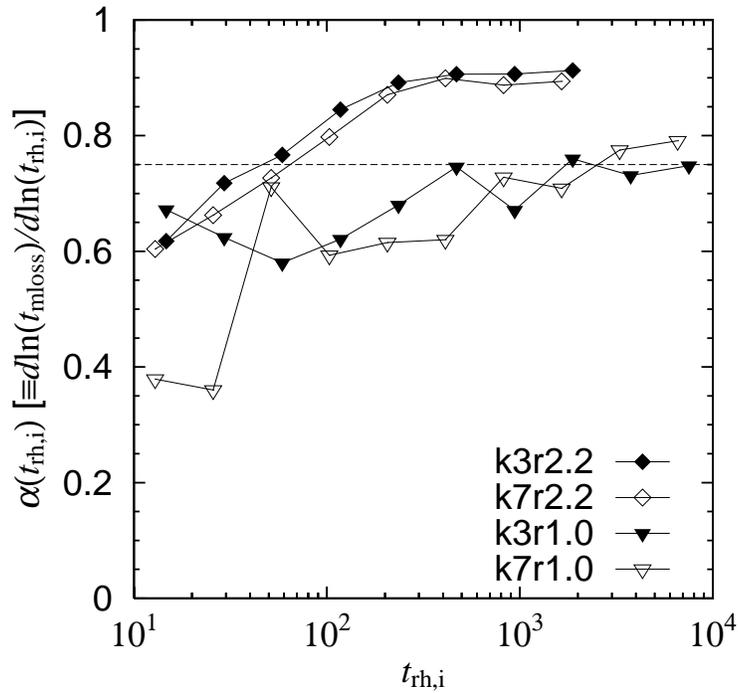}
\end{center}
\caption{Logarithmic slope $\alpha (t_{\rm rh,i})$[$\equiv d \ln (t_{\rm mloss}) / d \ln (t_{\rm rh,i})$]
as a function of the initial half-mass relaxation timescale for models k3r1.0,
k3r2.2, k7r1.0 and k7r2.2.
The dashed line indicates $\alpha (t_{\rm rh,i}) = 0.75$.}
\label{fig:fig8}
\end{figure}

We investigate the dependence of the mass loss timescale on initial
concentration of the cluster profile with $W_{0}=3$ and $7$ King profiles.
Figure \ref{fig:fig7} shows the mass loss timescale,
$t_{\rm mloss}$, as a function of the initial half-mass relaxation time,
$t_{\rm rh,i}$, for models k3r1.0, k3r2.2, k7r1.0 and k7r2.2.
In the models k7r1.0 and k7r2.2, the initial half-mass relaxation
timescale is estimated as
$t_{\rm rh,i} = 420 \times (N/8192)$.
The initial half-mass radii in the standard unit are $r_{\rm h,i}=0.77$,
for $W_{0}=7$.
As seen in figure \ref{fig:fig7}, the differences between models
k3r1.0 and k7r1.0 and between models k3r2.2 and k7r2.2 are small, which means
the mass loss timescale does not depend so much on the concentration of the
cluster.
Figure \ref{fig:fig8} shows the logarithmic slope,
$\alpha(t_{\rm rh,i})$ [ $\equiv d \ln (t_{\rm mloss}) / d \ln (t_{\rm rh,i})$ ].

\section{Discussion}

\subsection{Mass Loss Timescale for Small $N$}
\begin{figure}
\begin{center}
\FigureFile(120mm,75mm){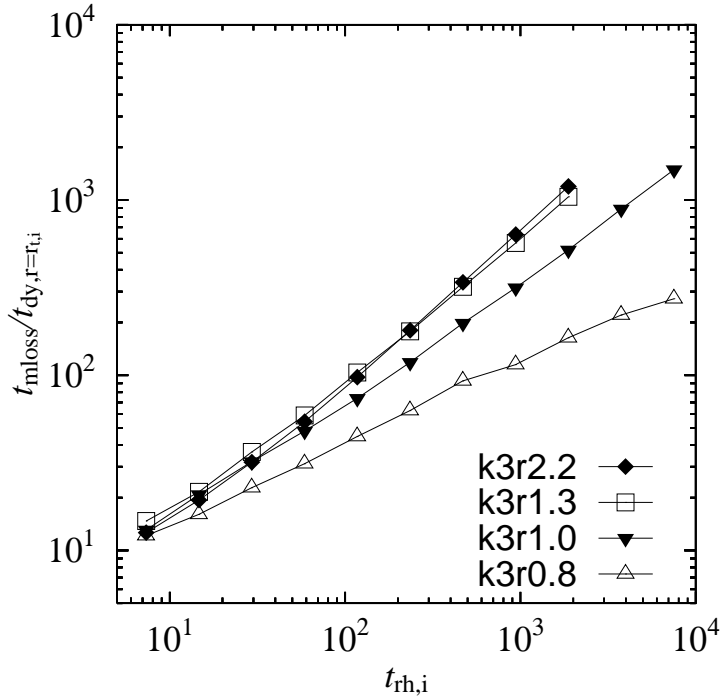}
\end{center}
\caption{Mass loss timescale scaled by the averaged dynamical time within the
tidal radius as a function of the initial half-mass relaxation times
for $W_{0}=3$ King cluster models.}
\label{fig:fig9}
\end{figure}

We discuss on the small $t_{\rm rh,i}$ (or small $N$) limit of the mass
loss timescale, $t_{\rm mloss}$, of clusters.
When $N$ is small, the half-mass relaxation time,
$t_{\rm rh,i}$, is sufficiently small compared with the escape time
delay, $t_{\rm e}$, and it is expected to dominate the total mass loss
timescale, $t_{\rm mloss}$.
In figure \ref{fig:fig9}, we show the mass loss timescale of
clusters scaled by the average dynamical time within the tidal radius,
$t_{\rm dy,r=r_{\rm t,i}}$, which is expressed as
\begin{equation}
t_{\rm dy,r=r_{\rm t,i}} = \frac{2{r_{\rm t,i}}^{3/2}}{G^{1/2}{M_{\rm i}}^{1/2}}
\end{equation}
and listed in table \ref{tab:series}, as a function of the initial half-mass
relaxation timescale for $W_{0}=3$ King cluster models.
The escape time delay is proportional to the average dynamical time
(FH).
In this figure, we don't plot that for model k3r4.5, since only this model
with $N=128$ and $256$ experiences core collapse before the half mass is lost.
As can be seen in figure \ref{fig:fig9}, all of the scaled mass
loss timescales, $t_{\rm rh,i}/t_{\rm dy,r=r_{\rm t,i}}$, approach to
$\sim 10$ around $t_{\rm rh,i} \sim 7$, which means that the mass loss
timescale is determined mainly by the averaged dynamical timescale or escape
time delay in small $t_{\rm rh,i}$ (or small $N$) limit.

\subsection{The Asymptotic Slope: $\alpha (t_{\rm rh,i}) \sim 0.9$}
In section 1, we conjecture that the mass loss timescale should be determined
only by the two-body relaxation timescale at its large limit.
Our simulation results show that when the external tidal field is not strong
(ex. model k3r2.2) the mass loss timescale is almost proportional to the
half-mass relaxation timescale.
However, its logarithmic slope seems to converge to slightly small value,
$\alpha (t_{\rm rh,i}) \sim 0.9$.
Here, we discuss origins of the small value $0.9$.

One possible answer is that the two-body relaxation timescale we used may not
correctly scale the real relaxation process occurred in the cluster.
Another possible answer is that our simulation time span may be still not long
enough to exclude the influence of the escape time delay on the mass loss
timescale.
Our simulation result and analytical estimate appear to favor the later
possibility.

\begin{figure}
\begin{center}
\FigureFile(120mm,75mm){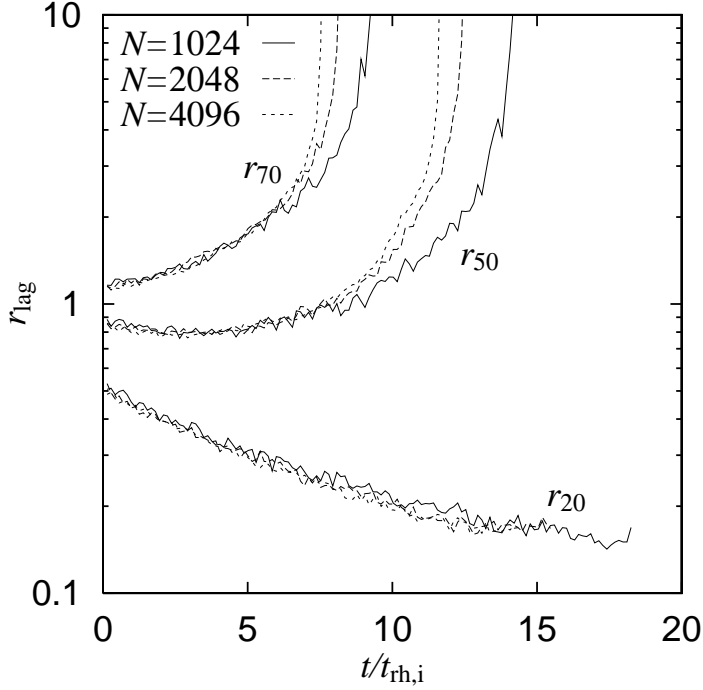}
\end{center}
\caption{Evolution of the Lagrangian radii containing 20, 50 and 70 \% of the
total mass as a function of time scaled by the initial half-mass relaxation
time.}
\label{fig:fig10}
\end{figure}

Figure \ref{fig:fig10} shows evolution of the Lagrangian radii as a function
of time scaled by the initial half-mass relaxation time, $t_{\rm rh,i}$ for
model k3r2.2 with $N=1024, 2048, 4096$.
We can see that the $20\%$ Lagrangian radii of clusters with different $N$ are
in good agreement, which indicates that the half-mass relaxation time is surely
a good measure of the internal relaxation process.

We estimate distribution of the escape time delay in model k7r2.2 using the
results of FH, which gave the distribution of escape time delay for fixed King
potential in $r_{\rm t,i} = r_{\rm kg}$ tidal field.
According to this paper, the fraction of the initial potential escapers
staying in a cluster, $P(t)$, is expressed by
\begin{equation}
P(t) = \sum_{\hat{E}} g_{\rm dis} (\hat{E}) \biggl\{ \Bigl[ 1 - f_{\rm non} (\hat{E}) \Bigr] (1+3.96 \omega t {\hat{E}}^2)^{-0.729} + f_{\rm non} (\hat{E}) \biggr\}, \label{eq:delay}
\end{equation}
where $\hat{E} = |(E - E_{\rm crit}) / E_{\rm crit}|$ and $E$ is the energy of
a potential escaper, $g_{\rm dis}(\hat{E})$ is the fraction of the initial
potential escapers, and $f_{\rm non}(\hat{E})$ is the fraction of non-escaper,
which can not escape from the cluster due to the regular orbits, in the initial
potential escapers.
We took $0.04$, $0.06$, $0.08$, $0.12$, $0.16$, and $0.24$ as representative
values of $\hat{E}$.

\begin{figure}
\begin{center}
\FigureFile(120mm,75mm){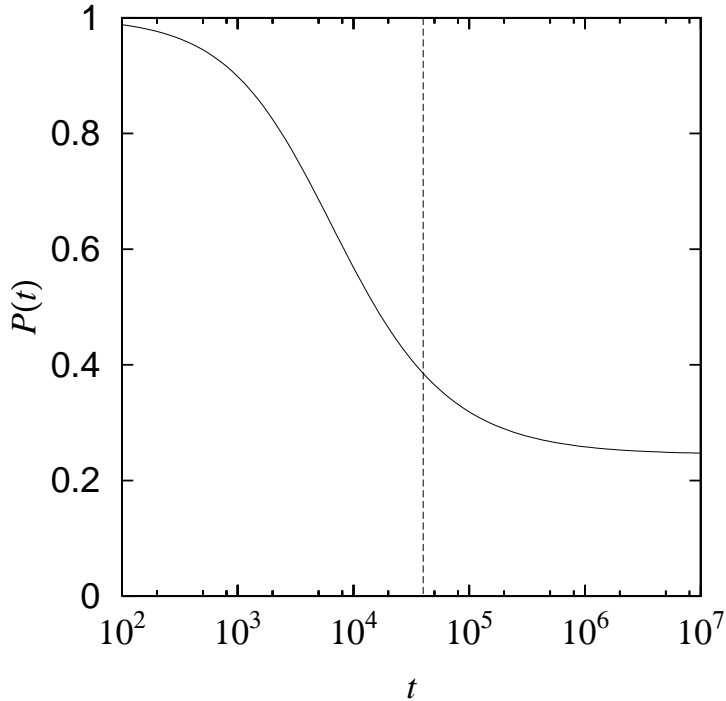}
\end{center}
\caption{Evolution of the fraction of the initial potential escapers based on
the results of FH in the case of k7r2.2 series.
Dashed line indicates $t=40000$, which is equal to the largest simulation span.}
\label{fig:fig11}
\end{figure}

Figure \ref{fig:fig11} shows the distribution of escape time delay equation
(\ref{eq:delay}).
In this figure we can see that at $t \sim 10^4$, which corresponds to the
maximum of our simulation span, only half of the initial potential escapers
have been depleted.
This means that the mass loss timescale, $t_{\rm mloss}$, of clusters are still
affected not only by the initial half-mass relaxation time,
$t_{\rm rh,i}$, but by the escape time delay, $t_{\rm e}$ within
our simulation span.
At least, 10 times simulation span may require to exclude the influence of the
escape time delay.

\section{Summary}
We investigate the evolution of star clusters in external tidal field by means
of $N$-body simulations.
We follows evolution of seven sets of clusters, whose dimensionless central
potential, $W_{0}$, of clusters and the strength of external tidal field are
different.

Our main conclusions are the following:
\begin{itemize}
\item[1.] The mass loss timescale depends on the strength of external tidal field.
\item[2.] The Logarithmic slopes,
$\alpha (t_{\rm rh,i})$ [ $\equiv d \ln ({t_{\rm mloss}}) / d \ln (t_{\rm rh,i})$],
also depend on the strength of external tidal field.
In model k3r1.0, whose run parameters are set to be the same as in B01, we can
reproduce the asymptotic power, $\alpha (t_{\rm rh,i}) \sim 0.75$.
However, when the tidal field is stronger (model k3r0.8) the slope is smaller
[$\alpha (t_{\rm rh,i}) \sim 0.4$], and when it is weaker (models k3r1.3,
k3r2.2, and k3r4.5) the slope is larger than
$\alpha (t_{\rm rh,i}) = 0.75$.
The slope $\alpha (t_{\rm rh,i})$ for the models in much weaker tidal
field (more k3r2.2 and k3r4.5) is seen to approach asymptotically to near
unity, but exactly $\alpha (t_{\rm rh,i}) \sim 0.9$.
\item[3.] The mass loss timescale is almost independent of the dimensionless central
potential, $W_{0}$.
\end{itemize}

\bigskip

We are grateful to Jun Makino and Holger Baumgardt for many helpful
discussions.
This research was partially supported by the Grants-in-Aid by the Japan Society
for the Promotion of Science (14740127) and by the Ministry of Education,
Science, Sports and Culture of Japan (16684002).

\appendix

\section{Reliability of Simulations}

\subsection{Force Accuracy}
\begin{figure}
\begin{center}
\FigureFile(120mm,75mm){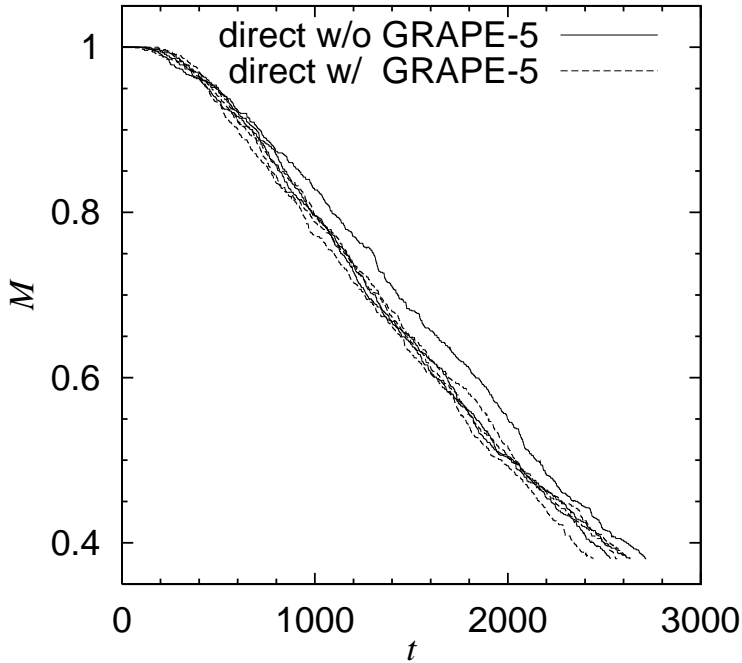}
\end{center}
\caption{Evolution of the total mass of cluster in model k3r2.2 ($N=1024$).
and solid lines are for runs with and without GRAPE-5, respectively.}
\label{fig:fig12}
\end{figure}

\begin{figure}
\begin{center}
\FigureFile(120mm,75mm){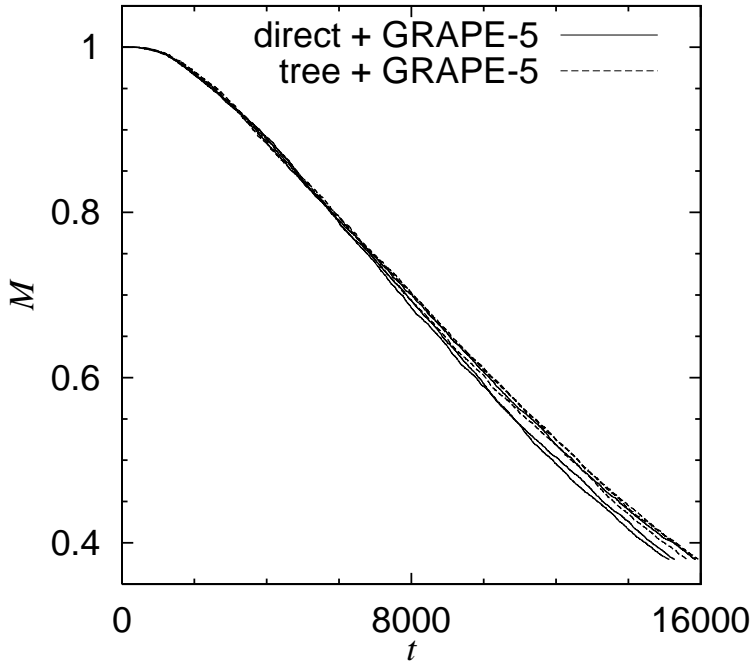}
\end{center}
\caption{Evolution of the total mass of clusters in model k3r2.2 ($N=8192$).
The solid and dashed lines indicates those obtained using direct summation and
the tree algorithm, respectively.
Three runs whose initial realization are different are performed for each
case.}
\label{fig:fig13}
\end{figure}

We use GRAPE-5 and the tree algorithm for the force calculations.
Both arise artificial errors in the calculated force.
When calculated with GRAPE-5 the force between two particles has a relative
error of $\sim 0.2$ percent, because of the low accuracy of internal
expressions in the GRAPE-5 chip (\cite{Makino1990}, \cite{Kawai2000}).
The force calculation by the tree algorithm also contains error of similar
magnitude, though it depends on the opening parameter.
However, such errors caused by GRAPE-5 and the tree algorithm may not
influence so much our results, because they are small compared to two-body
relaxation effects \citep{Hern1993}, which we want to handle correctly in our
simulations.
Figure \ref{fig:fig12} shows evolution of the total mass of clusters
when force calculations are done by the direct summation with and without
GRAPE-5, model k3r2.2 ($N=1024$).
The difference between runs with and without GRAPE-5 is smaller than its
run-to-run variation.
Figure \ref{fig:fig13} shows evolution of the total mass calculated
using direct summation and the tree algorithm ($\theta = 0.5$) in model
k3r2.2 ($N=8192$).
Both runs used GRAPE-5.
The difference of the results is sufficiently smaller than its run-to-run
variation.

\subsection{Accuracy of Time Integration}
\begin{figure}
\begin{center}
\FigureFile(120mm,75mm){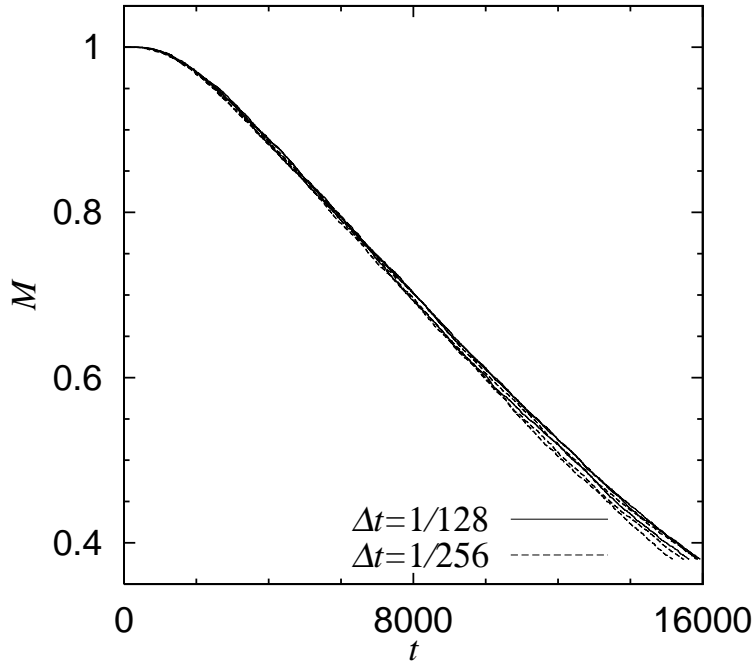}
\end{center}
\caption{Evolution of the total mass of clusters in model k3r2.2($N=8192$).
The solid and dashed lines indicates those obtained using timestep size
$\Delta t = 1/128$ and $\Delta t = 1/256$, respectively.
Three runs whose initial realization are different are performed for each
case.}
\label{fig:fig14}
\end{figure}

In our simulations, we use the leap-frog scheme, despite of the dependence of
forces on the velocities.
It is well-known that the nature of leap-frog scheme becomes worse in the
presence of this dependence.
However, figure \ref{fig:fig14} shows evolutions of the total
mass of clusters when $\Delta t = 1/128$ and $1/256$ are converged, which means
the leap-frog scheme with $\Delta t = 1/128$ does not influence our results.

\subsection{Potential Softening}
We use the potential softening for the force calculation.
One might suspect if the two-body relaxation doesn't happen in such smoothed
potential.
Using the potential softening, close encounter between particles is surely
suppressed.
However, energy of particles is changed dominantly by accumulation of distant
encounters, not by a few close encounters \citep{Spit1987}.
So, two-body relaxation should occur in the smoothed potential.

\begin{figure}
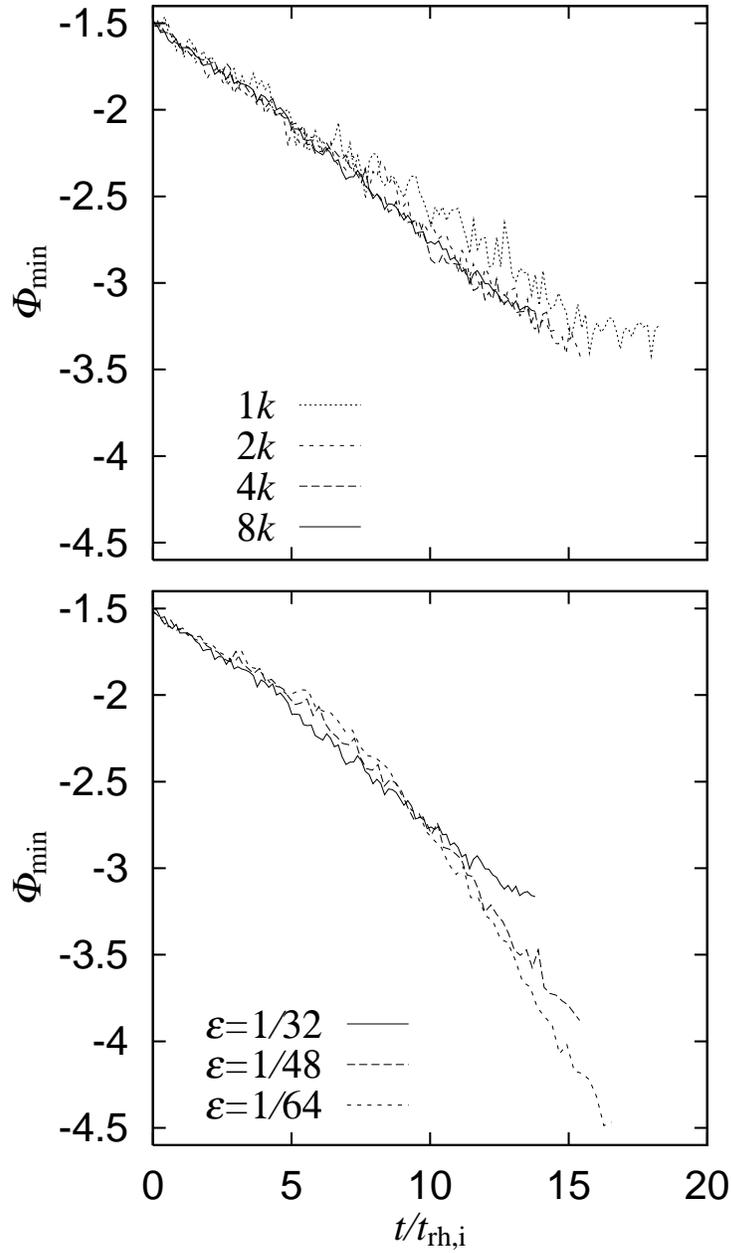

\begin{center}
\FigureFile(100mm,125mm){figure15.eps} \\
\end{center}
\caption{Evolution of the minimum potential in clusters as a function of
time scaled by the initial half-mass relaxation time in model k3r2.2 of
$N=1024, 2048, 4096, 8192$ and $\varepsilon = 1/32$ (top), and $N=8192$ and
$\varepsilon = 1/32, 1/48, 1/64$ (bottom).}
\label{fig:fig15}
\end{figure}

We check that the two-body relation happens with the softened potential.
The upper panel of figure \ref{fig:fig15} shows the evolutions of the
minimum potential in the clusters, $\Phi_{\rm min}$, as a
function of time scaled by the initial half-mass relaxation time,
$t_{\rm rh,i}$, in the clusters of model k3r2.2 of $N=1024, 2048, 4096, 8192$
and $\varepsilon = 1/32$.
The central potential decreases owing to the two-body relaxation effect.
As seen in the upper panel of figure \ref{fig:fig15}, these curves are in
good agreement and evolution of the central potential is scaled by the two-body
relaxation timescale.

The lower panel of figure \ref{fig:fig15} shows the evolutions of
$\Phi_{\rm min}$ in the clusters of model k3r2.2 of $N=8192$ and
$\varepsilon = 1/32, 1/48, 1/64$ as a function of timescale by the two-body
relaxation timescale [equation (\ref{eq:smoothed})].
These curves are also in good agreement, which indicates that even with a
softened potential, two-body relaxation occurs on the half-mass relaxation
timescale, described by equation (\ref{eq:smoothed}).

\end{document}